\documentclass[twocolumn,english,aps,pra,superscriptaddress,showpacs]{revtex4}
\usepackage{mathptmx}
\usepackage[T1]{fontenc}
\usepackage[latin1]{inputenc}
\usepackage{amsmath}
\usepackage{color}
\usepackage{graphicx}
\usepackage{amssymb}

\makeatletter
\usepackage{color}
\usepackage{graphics}

\renewcommand{\vec}{\mathbf}

\makeatother

\usepackage{babel}

\begin{document}

\title{Photo-detection using Bose-condensed atoms in a micro trap}

\author{S. Wallentowitz}

\affiliation{Facultad de Física, Pontificia Universidad Católica de Chile, Casilla
306, Santiago 22, Chile}

\author{A.B. Klimov}

\affiliation{Departamento de Física, Universidad de Guadalajara, Revolución 1500,
44410 Guadalajara, Jal., Mexico}

\begin{abstract}
A model of photo-detection using a Bose--Einstein condensate in an
atom-chip based micro trap is analyzed. Atoms absorb photons from
the incident light field, receive part of the photon momentum and
leave the trap potential. Upon counting of escaped atoms within predetermined
time intervals, the photon statistics of the incident light is mapped
onto the atom-count statistics. Whereas traditional photo-detection
theory treats the emission centers of photo electrons as distinguishable,
here the centers of escaping atoms are condensed and thus indistinguishable
atoms. From this an enhancement of the photon-number resolution as
compared to the commonly known counting formula is derived.
\end{abstract}

\date{May 24, 2008}


\maketitle

\section{Introduction}

The quantum theory of photo detection based on the absorption of photons
and emission of photo electrons represents one of the cornerstones
of quantum optics. It serves to obtain the statistics of emitted photo
electrons given the quantum statistics of the incident optical field.
Various approximations, as described by this theory, lead eventually
to the famous photo-counting formula of Mandel \citet{mandel2,kelley},
a quantum version of a previously known semi-classical Poissonian
formula \citet{mandel1,mandel3},\begin{equation}
P_{n}(t,t_{0})=\left\langle :\frac{[\eta_{D}\hat{I}(t,t_{0})]^{n}}{n!}e^{-\eta_{D}\hat{I}(t,t_{0})}:\right\rangle .\label{eq:mandel}\end{equation}
Here $:\;:$ denotes normal operator ordering, $\eta_{D}$ is the
quantum efficiency of the detector, and $\hat{I}(t,t_{0})$ is the
time-integrated light intensity incident on the detector's entrance
plane. 

This formula has the well-known limit of a purely Poissonian photo-electron
statistics, if the integration time $\tau=t-t_{0}$ is larger than
the coherence time of the incident optical field. This integration
time represents the response time of the detector system including
the connected electronics to amplify the generated photo currents.
Thus to observe the statistics of the optical field, one must ideally
have a fast detector and a large coherence time of the optical field
under study.

As already mentioned, to derive the Mandel formula, some approximations
have to be made. These approximations are perfectly justifiable for
a solid-state detector device that operates at not too low temperatures.
One crucial assumption is the distinguishability of the atoms emitting
the observable photo electrons. Another approximation is found to
consist in the perturbative calculus used to obtain joint probabilities
of photo-electron emissions. Together they lead to the Poissonian
operator form, rather independent of the underlying absorption dynamics.

Consider now a device that operates in a rather different regime,
that is, it may be cooled down to ultra cold temperatures in order
to behave more quantum than a typical solid-state photo detector.
For example, let us consider a cloud of magnetically trapped Bose-condensed
alkaline atoms \citet{bec} floating on the surface of a so-called
atom chip \citet{bec-on-atom-chip,atom-chip}. Atoms can now absorb
incident photons to receive part of the photon momentum, giving them
sufficient kinetic energy to escape from the trap, to subsequently
be detected, for instance by ionization.

As such a system is highly degenerate, the emission centers of escaping
atoms, i.e. the condensed atoms themselves, are not distinguishable.
Furthermore, a perturbative approach to calculate the emission probabilities
is hardly appropriate, as we may deal with Rabi cycles, where atoms
absorb and stimulatedly emit photons, thereby returning to the condensate.
Thus, the crucial approximations that led to the Mandel formula cannot
be applied and thus one may expect a rather different counting formula.
Such a counting formula connects the statistics of escaping atoms
to the statistics of the incident optical field. For the purpose of
unveiling the  different counting formula we study in the following
a model detector system using a Bose-condensed gas. Although, it serves
here merely for demonstrating the differences in the resulting counting
formula, we suppose that this system also possibly may be realizable
in current experiments. 

The paper is organized as follows: In Sec. \ref{sec:Photodetector-Model}
the model of the photo detector is introduced. The atom-counting statistics
is then derived in Sec. \ref{sec:Atom-Counting-Statistics}, followed
by a discussion of its features in Sec. \ref{sec:Discussion}. Finally,
a summary and conclusions are given in Sec. \ref{sec:Summary-and-Conclusions}.

\section{Photo detector Model\label{sec:Photodetector-Model}}

\subsection{Mechanism of photo-detection}

\begin{figure}
\includegraphics[width=0.48\textwidth]{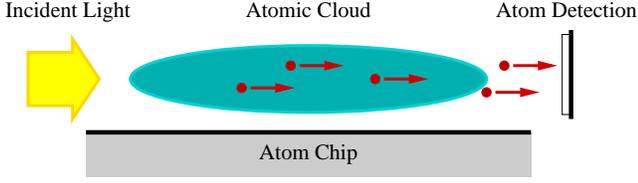}

\caption{Outline of a photo detector using trapped Bose-condensed atoms on
an atom chip. Atoms in the condensate absorb photons from the incident
light and escape from the trap, being then counted in the atom detector.}

\label{fig:outline}
\end{figure}

Let us assume that a cloud of bosonic atoms is magnetically trapped
in a micro trap implemented on an atom chip and being cooled well
below the condensation temperature, $T\ll T_{c}$. We suppose that
the trapping potential is highly elongated into one direction and
therefore approximate the system as being effectively one dimensional.
Furthermore, the atoms shall interact nearly resonant with a collimated
light field with incidence parallel to the elongated trap axis, see
Fig. \ref{fig:outline}. The overlap of the transverse mode structure
of the light with the transverse mode structure of the atomic cloud
shall be considered as a constant mode-matching parameter, that will
determine the coupling strength and thereby the efficiency of the
detector.

\begin{figure}
\includegraphics[width=0.48\textwidth]{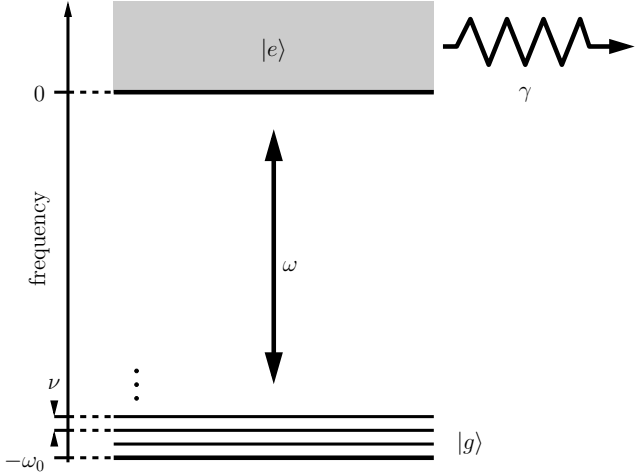}

\caption{Energy-level scheme of the atoms constituting the photo detector.
The zero-energy level is taken as the threshold, where the continuum
of unbound excited states starts. The ground-state level at frequency
$-\omega_{0}$ is superimposed by the vibrational sidebands of frequency
$\nu$. }

\label{fig:scheme} 
\end{figure}

The resonant electronic transition of the atoms shall be formed by
two levels, the lower level (ground state) being subject to the magnetic
trapping, whereas the upper level (excited state) being unaffected
by the trap. Thus, if all atoms start from the ground state, those
being excited by absorption of an incident photon may leave the trap
potential to be detected, e.g. by subsequent ionization. Some of the
excited atoms, however, will be de-excited by stimulated emission
and therefore will be subject to the trapping potential again. The
electronic level scheme including the sidebands generated by the trap
potential, and the loss of atoms from the trap is depicted in Fig.
\ref{fig:scheme}.

Given the atoms being Bose-condensed at the initial time $t_{0}$,
one may ask for the probability $P_{a}(t,t_{0})$ to observe $a$
atoms escaping from the trap in the time interval $[t_{0},t]$, during
which light is incident on the atomic probe. After this interval the
detector is reset, i.e. all atoms are cooled again into the condensate
mode (i.e. the lowest trap level), to start the counting again for
an identical time interval $\tau=t-t_{0}$. Thus $\tau$ will be the
analogue to the usual photo-detector integration time and the atom-counting
statistics $P_{a}$ will then be related, in a yet unknown way, to
the statistics of the incident optical field.

\subsection{Interaction with the incident optical field}

The Hamilton operator of the complete system including the detector
and the incident optical field can be decomposed into free and interaction
part as \begin{equation}
\hat{H}=\hat{H}_{0}+\hat{V},\label{hhv}\end{equation}
 where the free evolution is governed by \begin{equation}
\hat{H}_{0}=\hat{H}_{{\rm em}}+\hat{H}_{{\rm at}},\end{equation}
with $\hat{H}_{{\rm em}}$ being the Hamiltonian of the electromagnetic
field and $\hat{H}_{{\rm at}}$ being the Hamiltonian of the atoms. 

The electric field of the incident optical beam can be written as
a decomposition of monochromatic modes of wave vector $k$ and frequency
$ck$,\begin{equation}
\hat{E}(x)=\int dkE_{k}\hat{b}_{k}e^{ikx}+\mathrm{H.a.},\label{eq:E-expansion}\end{equation}
where $E_{k}$ denote the rms vacuum fluctuations of the electric-field
modes, $\hat{b}_{k}$ and $\hat{b}_{k}^{\dagger}$ are the bosonic
photon annihilation and creation operators, respectively. As the field
polarization is selected by the resonant atomic transition, only one
type of polarization is considered here. Using this expansion the
free Hamiltonian of the electromagnetic field thus becomes\begin{equation}
\hat{H}_{{\rm em}}=\int dkck\hat{b}_{k}^{\dagger}\hat{b}_{k}.\label{eq:ham-em}\end{equation}

The atomic system is described by the bosonic atom-field operators
$\hat{\Phi}_{i}(x)$, where $i=g,e$ denotes the electronic state,
and that satisfy the commutation relations\begin{equation}
[\hat{\Phi}_{i}(x),\hat{\Phi}_{j}^{\dagger}(x^{\prime})]=\delta_{ij}\delta(x-x^{\prime}).\end{equation}
The atomic Hamilton operator reads\begin{equation}
\hat{H}_{{\rm at}}=\sum_{i=g,e}\int dx\hat{\Phi}_{i}^{\dagger}(x)\left[-\frac{\hbar^{2}\partial_{x}^{2}}{2m}+\delta_{i,g}U(x)\right]\hat{\Phi}_{i}(x),\label{eq:ham-at}\end{equation}
 where the trap potential only acts in the electronic ground state
and reads\begin{equation}
U(x)=\frac{m\nu^{2}x^{2}}{2}-\hbar\omega_{eg},\end{equation}
with $\nu$ and $\omega_{eg}$ being the trap and electronic-transition
frequency, respectively, and $m$ is the atomic mass.

For the purpose of diagonalizing the free atomic Hamiltonian, we define
the Schrödinger eigen-modes:

\begin{equation}
\left[-\frac{\hbar^{2}\partial_{x}^{2}}{2m}+\delta_{i,g}U(x)\right]\phi_{i,n}(x)=\hbar\omega_{i,n}\phi_{i,n}(x).\end{equation}
Thus $\phi_{g,n}(x)$ with discrete $n=0,1,\ldots$ and eigen-frequencies\begin{equation}
\omega_{g,n}=n\nu-\omega_{0},\qquad(\omega_{0}=\omega_{eg}-{\textstyle \frac{\nu}{2}}),\end{equation}
are the harmonic-oscillator eigen-modes corresponding to a trapped
atom in its electronic ground state. The modes $\phi_{e,k}(x)\propto\exp(ikx)$
with continuous $k\in[-\infty,\infty]$ and eigen-frequencies \begin{equation}
\omega_{e,k}=\frac{\hbar k^{2}}{2m}.\end{equation}
are plane waves corresponding to a free atom in its electronic excited
state. These modes form two independent orthonormal sets and obey
the standard completeness relations \begin{eqnarray}
\sum_{n}\phi_{g,n}^{\ast}(x)\phi_{g,n}(x^{\prime}) & = & \delta(x-x^{\prime}),\\
\int dk\phi_{e,k}^{\ast}(x)\phi_{e,k}(x^{\prime}) & = & \delta(x-x^{\prime}).\end{eqnarray}

Each electronic component of the quantized atomic field can now be
expanded as

\begin{equation}
\hat{\Phi}_{g}(x)=\sum_{n}\hat{g}_{n}\phi_{g,n}(x),\qquad\hat{\Phi}_{e}(x)=\int dk\hat{e}_{k}\phi_{e,k}(x),\label{eq:Phi-expansion}\end{equation}
where the operators $\hat{g}_{n}$ and $\hat{e}_{k}$, each satisfy
again the bosonic commutation relations,\begin{equation}
[\hat{g}_{n},\hat{g}_{m}^{\dagger}]=\delta_{nm},\quad[\hat{e}_{k},\hat{e}_{k^{\prime}}^{\dagger}]=\delta(k-k^{\prime}).\end{equation}
Using the expansion (\ref{eq:Phi-expansion}) the free atomic Hamiltonian
(\ref{eq:ham-at}) reduces to \begin{equation}
\hat{H}_{{\rm at}}=\sum_{n}\hbar\omega_{g,n}\hat{g}_{n}^{\dagger}\hat{g}_{n}+\int dk\hbar\omega_{e,k}\hat{e}_{k}^{\dagger}\hat{e}_{k}.\end{equation}

The interaction between the atoms and the optical field reads in dipole
approximation\begin{equation}
\hat{V}=-d\kappa_{\perp}\int dx\sum_{i,j}\hat{\Phi}_{i}^{\dagger}(x)\hat{E}(x)\hat{\Phi}_{j}(x),\end{equation}
where $d$ is the transition dipole moment of the atoms and $\kappa_{\perp}$
is the matching between the transverse modes of electromagnetic and
atomic field. Using the expansions of electric and atomic fields,
cf. Eqs (\ref{eq:E-expansion}) and (\ref{eq:Phi-expansion}), respectively,
this interaction can be rewritten in optical rotating-wave approximation
as\begin{equation}
\hat{V}=-\sum_{n}\int dk^{\prime}\int dk\hbar\Omega_{k}\hat{e}_{k^{\prime}}^{\dagger}\hat{g}_{n}\hat{b}_{k}\underline{\phi}_{g,n}(k^{\prime}-k)+{\rm H.a.},\end{equation}
where the (vacuum) Rabi frequency has been defined as \begin{equation}
\Omega_{k}=\frac{dE_{k}\kappa_{\perp}}{\hbar},\end{equation}
and the Fourier transforms of the trap modes are defined as\begin{equation}
\underline{\phi}_{g,n}(k)=\frac{1}{\sqrt{2\pi}}\int\text{d}xe^{-ikx}\phi_{g,n}(x).\end{equation}

Thus the absorption of a photon of wave vector $k$ transforms a ground-state
atom in trap level $n$ into an excited-state atom with a superposition
of wave vectors $k^{\prime}$ given by $\underline{\phi}_{g,n}(k^{\prime}-k)$.
We may therefore define the annihilation operator of an excited wave
packet, created from trap level $n$ by absorption of a photon of
wave vector $k$:\begin{equation}
\hat{e}_{k,n}=\int dk^{\prime}\hat{e}_{k^{\prime}}\underline{\phi}_{g,n}^{\ast}(k^{\prime}-k).\label{eq:exc-wavepacket-op}\end{equation}
This relation can be inverted to obtain all operators $\hat{e}_{k}$
from the set of operators $\hat{e}_{k_{0},n}$ ($n=0,1,2,\ldots$)
for a specific wave vector $k_{0}$: \begin{equation}
\hat{e}_{k}=\sum_{n=0}^{\infty}\underline{\phi}_{g,n}(k-k_{0})\hat{e}_{k_{0},n}.\end{equation}
The excited-wavepacket operators satisfy the following commutation
relations\begin{equation}
[\hat{e}_{k,n},\hat{e}_{k^{\prime},n^{\prime}}^{\dagger}]=\int\text{d}x\phi_{g,n}^{\ast}(x)e^{i(k^{\prime}-k)x}\phi_{g,n^{\prime}}(x),\end{equation}
and in particular $[\hat{e}_{k,n},\hat{e}_{k,n}^{\dagger}]=1$.  Using
these wave packet operators, the interaction Hamiltonian can be simplified
to\begin{equation}
\hat{V}=-\sum_{n}\int dk\hbar\Omega_{k}\hat{e}_{k,n}^{\dagger}\hat{g}_{n}\hat{b}_{k}+{\rm H.a.}\label{eq:V-w-int}\end{equation}

\subsection{Single-mode approximation}

If we start from a Bose-condensed gas with all atoms being in the
lowest trap level $n=0$, a cycled electronic transition will preferably
lead again to the lowest trap level by bosonic enhancement. We may
thus approximate the ground-state levels by a single mode, corresponding
to the lowest trap level and may simplify the interaction Hamiltonian
(\ref{eq:V-w-int}) to\begin{equation}
\hat{V}=-\int dk\hbar\Omega_{k}\hat{e}_{k,0}^{\dagger}\hat{g}_{0}\hat{b}_{k}+{\rm H.a.}\end{equation}
Furthermore, we suppose that the incident optical field is quasi monochromatic
with wave vector $k_{0}$, so that only the photon operator $\hat{b}_{k_{0}}$has
to be kept. Using thus the definitions $\hat{b}_{k_{0}}\to\hat{b}$,
$\hat{g}_{0}\to\hat{g}$, $\hat{e}_{k_{0},0}\to\hat{e}$, and $\Omega_{k_{0}}\to\Omega$
the interaction further simplifies to\begin{equation}
\hat{V}=-\hbar\Omega\hat{e}^{\dagger}\hat{g}\hat{b}+{\rm H.a.}\label{eq:ham-int2}\end{equation}

The free atomic Hamiltonian, on the other hand, can be written in
this single-mode approximation as\begin{equation}
\hat{H}_{{\rm at}}=\hbar\overline{\omega}_{e}\hat{e}^{\dagger}\hat{e}-\hbar\omega_{0}\hat{g}^{\dagger}\hat{g}\label{eq:ham-at2}\end{equation}
with the average frequency of the excited wave packet being determined
by the wave vector of the absorbed photon and the momentum spread
of the ground-state trap level:\begin{equation}
\overline{\omega}_{e}=\int dk\omega_{e,k}|\underline{\phi}_{g,0}(k-k_{0})|^{2}=\frac{\nu}{4}+\frac{\hbar k_{0}^{2}}{2m}.\end{equation}
We note that the performed single-mode approximation neglects the
dispersion of the excited wave packet, being now considered as a propagating
plane wave.

The effective transition frequency between the relevant two atomic
levels becomes now\begin{equation}
\omega_{eg}^{\prime}=\overline{\omega}_{e}+\omega_{0}=\omega_{eg}+\frac{\hbar k_{0}^{2}}{2m}-\frac{\nu}{4}.\end{equation}
In resonance this transition frequency is compensated for by the frequency
$ck_{0}$ of the optical field, from which we obtain the resonance
condition\begin{equation}
ck_{0}\approx\omega_{eg}-\frac{\nu}{4}.\end{equation}

\subsection{Loss mechanism}

The spatio-temporal mode of the excited wave-packet in the single-mode
approximation is obtained as\begin{equation}
\phi_{e}(x,t)=e^{i(k_{0}x-\overline{\omega}_{e}t)}\phi_{g,0}^{\ast}(x).\end{equation}
It corresponds to a motion into the direction of the wave vector of
the previously absorbed photon with the group velocity $v_{g}=\overline{\omega}_{e}/k_{0}$
given by \begin{equation}
v_{g}=\frac{\hbar k_{0}}{2m}\left[1+\left(2\eta\right)^{-2}\right],\end{equation}
where $\eta=k_{0}\delta x_{0}$ is the Lamb--Dicke parameter with
$\delta x_{0}=\sqrt{\hbar/(2m\nu)}$ being the rms position spread
of the trap ground level. For a typical magnetic trap potential the
weak binding regime applies, where the Lamb--Dicke parameter is $\eta\gg1$.
Thus the conservation of momentum is approximately granted and the
excited atom compensates for the momentum of the absorbed photon.

If the excited wave packet has moved over a distance $\sim\delta x_{0}$
it may no longer be recycled into the electronic ground state by stimulated
emission of a photon, as the corresponding spatial overlap will be
close to zero. It thus has escaped from the trap. The time of flight
for this to happen is given by $\tau=\delta x_{0}/v_{g}$ and the
corresponding rate for this to happen, $\gamma=2\pi\tau^{-1}$, is
therefore obtained as\begin{equation}
\gamma=\frac{\hbar k_{0}^{2}}{m}\frac{\pi}{\eta}\left[1+(2\eta)^{-2}\right]\approx\frac{\pi\hbar k_{0}^{2}}{m\eta}.\end{equation}

The escape of atoms and their subsequent detection may be modeled
as an incoherent loss of atoms, described by the master equation\begin{equation}
\partial_{t}\hat{\varrho}=\frac{1}{i\hbar}[\hat{H},\hat{\varrho}]+\left.\partial_{t}\hat{\varrho}\right|_{{\rm esc}},\label{eq:master1}\end{equation}
where the atom loss is modeled by the Lindblad-form part\begin{equation}
\left.\partial_{t}\hat{\varrho}\right|_{{\rm esc}}=\gamma\left(\hat{e}\hat{\varrho}\hat{e}^{\dagger}-\frac{1}{2}\{\hat{e}^{\dagger}\hat{e},\hat{\varrho}\}\right).\end{equation}
The master equation (\ref{eq:master1}) can be written in the form\begin{equation}
\partial_{t}\hat{\varrho}=\frac{1}{i\hbar}\left(\hat{H}_{{\rm eff}}\hat{\varrho}-\hat{\varrho}\hat{H}_{{\rm eff}}^{\dagger}\right)+\gamma\hat{e}\hat{\varrho}\hat{e}^{\dagger},\end{equation}
where according to Eqs (\ref{eq:ham-em}), (\ref{eq:ham-int2}), and
(\ref{eq:ham-at2}) the non-Hermitean effective Hamilton operator
reads \begin{eqnarray}
\hat{H}_{{\rm eff}} & = & \hbar ck_{0}\hat{b}^{\dagger}\hat{b}+\hbar\overline{\omega}_{e}\hat{e}^{\dagger}\hat{e}-\hbar\omega_{0}\hat{g}^{\dagger}\hat{g}\label{eq:ham-eff}\\
 &  & -\hbar|\Omega|\left(\hat{e}^{\dagger}\hat{g}\hat{b}e^{i\varphi}+\hat{g}^{\dagger}\hat{e}\hat{b}^{\dagger}e^{-i\varphi}\right)-\frac{i\hbar\gamma}{2}\hat{e}^{\dagger}\hat{e},\nonumber \end{eqnarray}
with $\Omega=|\Omega|e^{i\varphi}$.

\subsection{Atomic pseudo spin}

For the atomic quantum fields we may define the pseudo spin operator
$\hat{\vec{S}}$ with components ($\hat{S}_{\pm}=\hat{S}_{x}\pm i\hat{S}_{y}$)

\begin{eqnarray}
\hat{S}_{+} & = & \hat{e}^{\dagger}\hat{g}e^{i\varphi},\qquad\hat{S}_{-}=(\hat{S}_{+})^{\dagger},\label{eq:S+-def}\\
\hat{S}_{z} & = & {\textstyle \frac{1}{2}}(\hat{e}^{\dagger}\hat{e}-\hat{g}^{\dagger}\hat{g}),\label{eq:Sz-def}\end{eqnarray}
and\begin{equation}
\hat{S}^{2}=\frac{\hat{A}}{2}\left(\frac{\hat{A}}{2}+1\right),\label{eq:S2-def}\end{equation}
where the atom-number operator is defined as\begin{equation}
\hat{A}=\hat{e}^{\dagger}\hat{e}+\hat{g}^{\dagger}\hat{g}.\label{eq:A-op}\end{equation}
The operators $\hat{S}_{\pm,z}$ satisfy the standard su(2) commutation
relations, $[\hat{S}_{+},\hat{S}_{-}]=2\hat{S}_{z}$ and $[\hat{S}_{z},\hat{S}_{\pm}]=\pm\hat{S}_{\pm}$.
The total number of excitations -- excited atoms plus photons -- is
given by the operator\begin{equation}
\hat{N}=\hat{e}^{\dagger}\hat{e}+\hat{b}^{\dagger}\hat{b}=\hat{S}_{z}+\frac{\hat{A}}{2}+\hat{b}^{\dagger}\hat{b}.\label{eq:N-def}\end{equation}

The atomic system can now be described in the basis of Dicke states
\citet{dicke} \begin{equation}
|A,A_{e}\rangle_{{\rm at}}=|A-A_{e}\rangle_{g}\otimes|A_{e}\rangle_{e},\end{equation}
where $A$ is the total number of atoms, i.e. $\hat{A}|A,A_{e}\rangle_{{\rm at}}=A|A,A_{e}\rangle_{{\rm at}}$,
and $A_{e}$ is the number of excited atoms. The corresponding basis
states for the total system can then be written as\begin{equation}
|A,N,n\rangle=|{\textstyle A},N-n\rangle_{{\rm at}}\otimes|n\rangle_{{\rm em}},\label{eq:basis-states}\end{equation}
where $N$ is the total number of excitations, i.e. $\hat{N}|A,N,n\rangle=N|A,N,n\rangle$,
and $|n\rangle_{{\rm em}}$ is a photon-number state. 

Using the definitions of the spin operators (\ref{eq:S+-def}-\ref{eq:N-def}),
the effective Hamiltonian (\ref{eq:ham-eff}) can be rewritten as

\begin{equation}
\hat{H}_{{\rm eff}}=\hat{H}_{0}^{\prime}-\hbar\Delta^{\prime}\hat{S}_{z}-\frac{i\hbar\gamma}{4}\hat{A}-\hbar|\Omega|\left(\hat{S}_{+}\hat{b}+\hat{S}_{-}\hat{b}^{\dagger}\right),\end{equation}
where for notational simplicity we defined $\Delta^{\prime}=\Delta+i\gamma/2$
with $\Delta=ck_{0}-\omega_{eg}^{\prime}$ being the detuning from
resonance, and the free Hamiltonian is identified as\begin{equation}
\hat{H}_{0}^{\prime}=\hbar ck_{0}\hat{N}+\frac{\hbar(\overline{\omega}_{e}-\omega_{0})}{2}\hat{A}.\end{equation}

As this free part commutes with the remainder of the effective Hamiltonian,
we may transform into the interaction picture with respect to $\hat{H}_{0}^{\prime}$,
to obtain the master equation in the interaction picture\begin{equation}
\partial_{t}\hat{\varrho}=\frac{1}{i\hbar}\left(\hat{H}_{{\rm eff}}\hat{\varrho}-\hat{\varrho}\hat{H}_{{\rm eff}}^{\dagger}\right)+\gamma\hat{e}\hat{\varrho}\hat{e}^{\dagger},\label{eq:master}\end{equation}
where the transformed effective Hamiltonian becomes\begin{equation}
\hat{H}_{{\rm eff}}=-\hbar\Delta^{\prime}\hat{S}_{z}-\frac{i\hbar\gamma}{4}\hat{A}-\hbar|\Omega|\left(\hat{S}_{+}\hat{b}+\hat{S}_{-}\hat{b}^{\dagger}\right).\label{eq:H-eff}\end{equation}
For notational convenience we omitted here any indication of being
in the interaction picture. 

In the master equation (\ref{eq:master}), the last term describes
the escape of an excited-wavepacket atom from the trap. The responsible
operator $\hat{e}$, that annihilates one such excited atom from the
system, can be written in the basis of the states (\ref{eq:basis-states})
as \begin{eqnarray}
\hat{e} & = & \sum_{A,N,n}\sqrt{N-n}|A-1,N-1,n\rangle\langle A,N,n|.\label{eq:e-op}\end{eqnarray}

\subsection{Limit of large number of atoms}

Let us now consider the action of the operators appearing in the effective
Hamiltonian (\ref{eq:H-eff}) on the basis states (\ref{eq:basis-states}).
Firstly, the actions of the operators $\hat{S}_{+}\hat{b}$ and $\hat{S}_{z}$
on a state $|A,N,n\rangle$ are:\begin{eqnarray}
\hat{S}_{+}\hat{b}|A,N,n\rangle & = & \sqrt{(N-n+1)(A-N+n)n}\label{eq:S+-action}\\
 &  & \quad\times|A,N,n-1\rangle,\nonumber \\
\hat{S}_{z}|A,N,n\rangle & = & (N-n-\frac{A}{2})|A,N,n\rangle.\label{eq:Sz-action}\end{eqnarray}
Defining a new spin operator $\hat{\vec{L}}$ with its components
being defined by the actions \begin{eqnarray}
\hat{L}_{+}|A,N,n\rangle & = & \sqrt{(N-n)(n+1)}|A,N,n+1\rangle,\\
\hat{L}_{-}|A,N,n\rangle & = & \sqrt{(N-n+1)n}|A,N,n-1\rangle,\\
\hat{L}_{z}|A,N,n\rangle & = & (n-\frac{N}{2})|A,N,n\rangle,\end{eqnarray}
and obeying the usual angular-momentum commutation relations, Eqs
(\ref{eq:S+-action}) and (\ref{eq:Sz-action}) can be rewritten as\begin{eqnarray}
\hat{S}_{+}\hat{b}|A,N,n\rangle & = & \hat{L}_{-}\left(\hat{A}-\frac{\hat{N}}{2}+\hat{L}_{z}\right)^{\frac{1}{2}}|A,N,n\rangle,\\
\hat{S}_{z}|A,N,n\rangle & = & \left(\frac{\hat{N}-\hat{A}}{2}-\hat{L}_{z}\right)|A,N,n\rangle.\end{eqnarray}
Thus accordingly we may replace the operators in the effective Hamiltonian
(\ref{eq:H-eff}) to obtain\begin{eqnarray}
\hat{H}_{\mathrm{eff}} & = & -\hbar\Delta^{\prime}\left(\frac{\hat{N}}{2}-\hat{L}_{z}\right)+\hbar\Delta\frac{\hat{A}}{2}\nonumber \\
 &  & -\hbar|\Omega_{0}|\left[\hat{L}_{-}\sqrt{\hat{M}+\hat{L}_{z}-{\textstyle \frac{1}{2}}}+{\rm H.a.}\right].\label{eq:heff-L}\end{eqnarray}
where the operator \begin{equation}
\hat{M}=\hat{A}-\frac{\hat{N}}{2}+\frac{1}{2},\end{equation}
has been introduced.

For a proper functioning of the detector we assume that the number
of atoms in the gas is much larger than the maximum number of photons
of the incident optical field. Thus the occupied eigenvalues of the
operator $\hat{M}$ are very large and consequently we may perform
an expansion over a small parameter being proportional to the inverse
atom number \citet{chumakov1,chumakov2}. Thus the interaction part
of the effective Hamiltonian is expanded as\begin{eqnarray}
\hat{L}_{-}\sqrt{\hat{M}+\hat{L}_{z}-{\textstyle \frac{1}{2}}} & = & \hat{L}_{-}\sqrt{\hat{M}}\left(1+\frac{\hat{L}_{z}-\frac{1}{2}}{\hat{M}}\right)^{\frac{1}{2}}\nonumber \\
 & = & \hat{L}_{-}\sqrt{\hat{M}}\left(1+\frac{\hat{L}_{z}-\frac{1}{2}}{2\hat{M}}+\ldots\right),\end{eqnarray}
where the expansion parameter $(\hat{L}_{z}-\frac{1}{2})/\hat{M}$
is chosen to cancel the first-order contribution in (\ref{eq:heff-L}),
obtaining the zero-order Hamiltonian as\begin{equation}
\hat{H}_{\mathrm{eff}}(t)\approx-\hbar\Delta^{\prime}\left(\frac{\hat{N}}{2}-\hat{L}_{z}\right)+\hbar\Delta\frac{\hat{A}}{2}-2\hbar|\Omega|\sqrt{\hat{M}}\hat{L}_{x}.\end{equation}

\section{Atom-Counting Statistics\label{sec:Atom-Counting-Statistics}}

\subsection{Counting statistics }

From the solution of the master equation we need to extract the probability
for $a$ atoms having escaped in the time interval $[t_{0},t]$, starting
at the initial time $t_{0}$ with a perfect Bose-condensed gas with
$A$ atoms in the trap ground state. Thus the initial state at time
$t_{0}$ can be written in the form\begin{equation}
\hat{\varrho}(t_{0})=|A,0\rangle_{{\rm at}}\langle A,0|\otimes\hat{\rho}_{{\rm em}}(t_{0}),\end{equation}
where $\hat{\rho}_{{\rm em}}(t_{0})$ is the density operator of the
incident optical field at the initial time $t_{0}$. The latter may
be expanded in photon-number states as\begin{equation}
\hat{\rho}_{{\rm em}}(t_{0})=\sum_{n,n^{\prime}}\rho_{n,n^{\prime}}(t_{0})|n\rangle_{{\rm em}}\langle n^{\prime}|,\end{equation}
so that the complete initial density operator can be written in the
basis states (\ref{eq:basis-states}) as\begin{equation}
\hat{\varrho}(t_{0})=\sum_{n,n^{\prime}}\rho_{n,n^{\prime}}(t_{0})|A,n,n\rangle\langle A,n^{\prime},n^{\prime}|.\label{eq:rho-ini}\end{equation}

Given the initial density operator (\ref{eq:rho-ini}), the formal
solution of the density operator at time $t\geq t_{0}$ can be obtained
from the master equation in the form\begin{eqnarray}
\hat{\varrho}(t,t_{0}) & = & \sum_{a=0}^{\infty}\hat{\varrho}_{a}(t,t_{0}),\end{eqnarray}
where $\hat{\varrho}_{a}(t,t_{0})$ is the (unnormalized) conditional
density operator corresponding to the history of the detector system
where in total $a$ atoms have escaped in the time interval $[t_{0},t]$.
The norm of this conditional density operator is the probability for
this history to occur, which is the desired probability to count $a$
atoms escaping from the trap:\begin{equation}
P_{a}(t,t_{0})={\rm Tr}\left[\hat{\varrho}_{a}(t,t_{0})\right].\end{equation}

The conditional density operator $\hat{\varrho}_{a}(t,t_{0})$ is
itself a sum of all possible histories where $a$ atoms escape in
such a way, that the $j$th atom escapes at time $t_{j}\in[t_{0},t]$,
where $j=1,\ldots,a$, \begin{equation}
\hat{\varrho}_{a}(t,t_{0})=\int_{t_{0}}^{t}dt_{a}\ldots\int_{t_{0}}^{t_{2}}dt_{1}\hat{\varrho}_{a}(t,t_{0};t_{a},\ldots,t_{1}).\label{eq:cond-dens-op}\end{equation}
The norm of the conditional density operator on the rhs is the joint
probability density for $a$ atoms to escape from the trap at times
$t_{1},\ldots,t_{a}$:\begin{equation}
p_{a}(t,t_{0};t_{a},\ldots,t_{1})={\rm Tr}\left[\hat{\varrho}_{a}(t,t_{0};t_{a},\ldots,t_{1})\right].\label{eq:prob-dens}\end{equation}
Thus the required counting statistics is obtained as\begin{equation}
P_{a}(t,t_{0})=\int_{t_{0}}^{t}dt_{a}\ldots\int_{t_{0}}^{t_{2}}dt_{1}p_{a}(t,t_{0};t_{a},\ldots,t_{1}).\label{eq:Pa-def}\end{equation}

\subsection{Quantum trajectories}

Given the initial density operator (\ref{eq:rho-ini}), the conditional
density operator $\hat{\varrho}_{a}(t,t_{0};t_{a},\ldots,t_{1})$
is given by the quantum trajectory \citet{qt1,qt2,qt3} \begin{widetext}

\begin{equation}
\hat{\varrho}_{a}(t,t_{0};t_{a},\ldots,t_{1})=\hat{\mathcal{N}}(t-t_{a})\hat{\mathcal{J}}\hat{\mathcal{N}}(t_{a}-t_{a-1})\hat{\mathcal{J}}\ldots\ldots\hat{\mathcal{J}}\hat{\mathcal{N}}(t_{1}-t_{0})\hat{\varrho}(t_{0}).\label{eq:trajectory}\end{equation}
\end{widetext}It is a non-unitary evolution $\hat{\mathcal{N}}$
intermittent by so-called jump operators $\hat{\mathcal{J}}$ that
describe the escape of a single atom from the trap. The super operator
of the non-unitary evolution is defined as\begin{eqnarray}
\hat{\mathcal{N}}(t)\hat{\varrho} & = & \hat{U}_{\mathrm{eff}}(t)\hat{\varrho}\hat{U}_{\mathrm{eff}}^{\dagger}(t),\label{eq:N-op}\end{eqnarray}
 the effective non-unitary evolution operator being \begin{equation}
\hat{U}_{{\rm eff}}(t)=\exp\left(-\frac{it}{\hbar}\hat{H}_{\mathrm{eff}}\right),\label{eq:Ueff}\end{equation}
and the escape of an atom is described by the super operator\begin{equation}
\hat{\mathcal{J}}\hat{\varrho}=\gamma\hat{e}\hat{\varrho}\hat{e}^{\dagger}.\label{eq:J-op}\end{equation}

Thus the joint probability density (\ref{eq:prob-dens}) can be written
using Eqs (\ref{eq:trajectory})-(\ref{eq:J-op}) and the initial
state (\ref{eq:rho-ini}) as \begin{widetext}

\begin{equation}
p_{a}(t,t_{0};t_{a},\ldots,t_{1})=\sum_{n,n^{\prime}}\rho_{n,n^{\prime}}(t_{0})\langle\Phi(t,t_{0};t_{a},\ldots,t_{1}|n^{\prime})|\Phi(t,t_{0};t_{a},\ldots,t_{1}|n)\rangle,\label{eq:prob-dens2}\end{equation}
where the (unnormalized) quantum-trajectory states starting with the
initial state $|A,n,n\rangle$ are\begin{equation}
|\Phi(t,t_{0};t_{a},\ldots,t_{1}|n)\rangle=\gamma^{a/2}\hat{U}_{{\rm eff}}(t-t_{a})\hat{e}\ldots\hat{e}\hat{U}_{{\rm eff}}(t_{1}-t_{0})|A,n,n\rangle.\label{eq:trajectory1}\end{equation}
Using the representation of the operator $\hat{e}$ in the basis states,
Eq. (\ref{eq:e-op}), this state vector can be rewritten as 

\begin{eqnarray}
|\Phi(t,t_{0};t_{a},\ldots,t_{1}|n)\rangle & = & \sum_{n_{a},\ldots,n_{1}}\hat{U}_{{\rm eff}}(t-t_{a})|A-a,n-a,n_{a}\rangle\nonumber \\
 & \times & \Psi_{n_{a},n_{a-1}}^{A-a+1,n-a+1}(t_{a}-t_{a-1})\ldots\Psi_{n_{2},n_{1}}^{A-1,n-1}(t_{2}-t_{1})\Psi_{n_{1},n}^{A,n}(t_{1}-t_{0}),\label{eq:trajectory2}\end{eqnarray}
\end{widetext}where the transition amplitudes are\begin{equation}
\Psi_{m,n}^{A,N}(t)=\sqrt{\gamma(N-m)}\langle A,N,m|\hat{U}_{{\rm eff}}(t)|A,N,n\rangle.\label{eq:trans-ampl}\end{equation}
Here we made use of the fact that the effective Hamiltonian preserves
both the atom number and the total number of excitations. As the trajectory
(\ref{eq:trajectory2}) has exactly $A-a$ atoms and $n-a$ remaining
excitations, in the sum of Eq. (\ref{eq:prob-dens2}) only terms with
$n=n^{\prime}$ contribute:\begin{equation}
p_{a}(t,t_{0};t_{a},\ldots,t_{1})=\sum_{n}P_{n}p_{a}(t,t_{0};t_{a},\ldots,t_{1}|n),\label{eq:prob-dens3}\end{equation}
where $P_{n}=\rho_{n,n}(t_{0})$ is the initial photon statistics
and the probability density conditioned on initially $n$ photons
is defined as\begin{equation}
p_{a}(t,t_{0};t_{a},\ldots,t_{1}|n)=\left\Vert |\Phi(t,t_{0};t_{a},\ldots,t_{1}|n)\rangle\right\Vert ^{2}.\label{eq:cond-prob-dens}\end{equation}

Thus the atom-counting statistics can be written as\begin{equation}
P_{a}(t,t_{0})=\sum_{n}P_{a}(t,t_{0}|n)P_{n},\end{equation}
where the conditional probability for $a$ atoms to escape in the
time interval $[t_{0},t]$ given that $n$ photons are present is
\begin{equation}
P_{a}(t,t_{0}|n)=\int_{t_{0}}^{t}dt_{a}\ldots\int_{t_{0}}^{t_{2}}dt_{1}p_{a}(t,t_{0};t_{a},\ldots,t_{1}|n).\label{eq:Pan}\end{equation}

\subsection{Over damped resonant regime}

The features of the dynamics of the absorption and stimulated emission
of photons and the loss of atoms from the trap, given a state with
$A$ atoms and $N$ excitations, depends on the saturation parameter
(cf. App. \ref{sec:Transition-amplitude})\begin{equation}
S_{A,N}=\frac{4|\Omega|^{2}M_{A,N}}{\Delta^{2}+(\gamma/2)^{2}},\end{equation}
where $\hat{M}|A,N,n\rangle=M_{A,N}|A,N,n\rangle$ with $M_{A,N}=A-(N-1)/2$.
Given perfect resonance of the incident monochromatic light field,
$\Delta=0$, for $S_{A,N}>1$ Rabi oscillations occur that involve
cycles of absorption and (stimulated) re-emission of photons until
an atom is lost from the trap. 

In the opposite over damped case, $S_{A,N}<1$, however, no cycling
transition is observed but photons are absorbed and their excitation
is removed from the system by an atom leaving the trap. In other words,
the recoil energy is much larger than the effective coupling energy
of the atoms with the photon field, \begin{equation}
\frac{(\hbar k_{0})^{2}}{2m}>\frac{\eta}{\pi}\left[2\hbar|\Omega|\sqrt{M_{A,N}}\right].\end{equation}
We recall that in our case the Lamb--Dicke parameter $\eta>1$. 

Thus the number of absorbed photons equals approximately the number
of lost atoms, which means that only the transition amplitudes\begin{equation}
\Psi_{m,n}^{A,n}(t)=\sqrt{\gamma(n-m)}\langle A,n,m|\hat{U}_{{\rm eff}}(t)|A,n,n\rangle\label{eq:trans-ampl2}\end{equation}
have to be considered, since only they start from initial states with
all excitation being photonic. From Eq. (\ref{eq:trajectory2}) it
becomes then clear that among these transition amplitudes we may further
consider only those where one photon has been absorbed, i.e.,\begin{equation}
\Psi_{n-1,n}^{A,n}(t)=\sqrt{\gamma}\langle A,n,n-1|\hat{U}_{{\rm eff}}(t)|A,n,n\rangle.\label{eq:trans-ampl3}\end{equation}

However, as these are approximations we must re-normalize correctly
the transition amplitudes to obtain statistically correct quantum
trajectories. Originally the normalization read\begin{equation}
\sum_{m=0}^{N}\int_{0}^{\infty}dt|\Psi_{m,n}^{A,N}(t)|^{2}=1,\end{equation}
meaning that starting from a state $|A,N,n\rangle$ the system eventually
will end up with certainty in one of the states $|A,N,m\rangle$ with
$m=0,\ldots,N$. Taking into account now only the transition amplitudes
(\ref{eq:trans-ampl3}), we must use the re-normalized transition
amplitude \begin{equation}
\Psi_{n-1,n}^{A,n}(t)\to\Psi_{A,n}(t)=\Psi_{n-1,n}^{A,n}(t)/\sqrt{P_{n-1,n}^{A,n}},\label{eq:trans-ampl4}\end{equation}
where the probability for the considered transition is defined as
\begin{equation}
P_{n-1,n}^{A,n}=\int_{0}^{\infty}dt|\Psi_{n-1,n}^{A,n}(t)|^{2}\simeq1.\label{eq:prob-trans}\end{equation}
In this way we obtain an atom waiting-time distribution\begin{equation}
w_{A,n}(t)=|\Psi_{A,n}(t)|^{2},\label{eq:w-def}\end{equation}
that is properly normalized:\begin{equation}
\int_{0}^{\infty}dtw_{A,n}(t)=1.\end{equation}

The quantum trajectory in the over damped regime is thus simplified
to\begin{eqnarray}
|\Phi(t,t_{0};t_{a},\ldots,t_{1}|n)\rangle & = & \hat{U}_{{\rm eff}}(t-t_{a})|A-a,n-a,n-a\rangle\nonumber \\
 & \times & \prod_{k=0}^{a-1}\Psi_{A-k,n-k}(t_{k+1}-t_{k}),\end{eqnarray}
so that the joint probability density (\ref{eq:cond-prob-dens}) becomes
\begin{eqnarray}
p_{a}(t,t_{0};t_{a},\ldots,t_{1}|n) & = & W_{A-a,n-a}(t-t_{a})\label{eq:cond-prob-dens2}\\
 & \times & \prod_{k=0}^{a-1}w_{A-k,n-k}(t_{k+1}-t_{k}),\nonumber \end{eqnarray}
where \begin{equation}
W_{A,n}(t)=\langle A,n,n|\hat{U}_{{\rm eff}}^{\dagger}(t)\hat{U}_{{\rm eff}}(t)|A,n,n\rangle\label{eq:W-def}\end{equation}
is the probability that no atom leaves the trap within the time interval
$t$ starting from the state $|A,n,n\rangle$. 

When approximating the transition amplitudes, cf. Eq. (\ref{eq:trans-ampl4}),
to obtain statistically correct quantum trajectories that lead to
a normalized atom-count statistics $P_{a}$, also the probability
(\ref{eq:W-def}) must be consistently approximated. This can be done
by using the general relation\begin{equation}
W_{A,n}(t)=1-\int_{0}^{t}dt^{\prime}w_{A,n}(t^{\prime}),\label{eq:no-escape-prob}\end{equation}
employing on the rhs the approximation (\ref{eq:w-def}).

Using the above results, the conditional probability for $a$ atoms
to leave the trap given that $n$ photons are present, Eq. (\ref{eq:Pan}),
becomes now\begin{eqnarray}
P_{a}(t,t_{0}|n) & = & \int_{t_{0}}^{t}dt_{a}\ldots\int_{t_{0}}^{t_{2}}dt_{1}W_{A-a,n-a}(t-t_{a})\nonumber \\
 &  & \times\prod_{k=0}^{a-1}w_{A-k,n-k}(t_{k+1}-t_{k}).\label{eq:Pa-def1}\end{eqnarray}
This convolution integral is expressed as the inverse Laplace transform
\begin{equation}
P_{a}(\tau|n)=\mathcal{L}^{-1}\left[\underline{W}_{A-a,n-a}(z)\prod_{k=0}^{a-1}\underline{w}_{A-k,n-k}(z)\right],\label{eq:product}\end{equation}
where $\tau=t-t_{0}$.

In the resonant case ($\Delta=0$), low saturation ($S_{A}\ll1$),
and assuming numbers of photons much lower than the atom number, $n\ll A$,
we obtain the atom waiting-time distribution as {[}cf. Eq. (\ref{eq:psi-approx})]\begin{equation}
w_{A,n}(t)\propto\sinh^{2}\left(\frac{\gamma t}{4}\sqrt{1-S_{A}}\right)e^{-\frac{\gamma t}{2}[n-(n-1)\sqrt{1-S_{A}}]},\label{eq:w-approx}\end{equation}
where the saturation parameter (\ref{eq:saturation1}) has been approximated
by $S_{A}=4\Omega^{2}A/(\gamma/2)^{2}$. This form shows a behavior
quite similar to over damped Rabi oscillations of a two-level system
interacting with a resonant laser field, with the saturation being
now dependent on the number of atoms.  The individual Laplace transforms
become then\begin{eqnarray}
 &  & \underline{w}_{A-k,n-k}(z)=\\
 &  & \quad\frac{\left(n-k\right)\left(n-k+q\right)\left(n-k+2q\right)}{\left(n-k+\tau_{0}z\right)\left(n-k+\tau_{0}z+q\right)\left(n-k+\tau_{0}z+2q\right)},\nonumber \end{eqnarray}
where we defined\begin{equation}
q=\frac{\sqrt{1-S_{A}}}{1-\sqrt{1-S_{A}}},\qquad\tau_{0}^{-1}=\frac{\gamma}{2}\left(1-\sqrt{1-S_{A}}\right).\label{eq:q-def}\end{equation}
As we deal with low saturation, $S_{A}\ll1$, the introduced parameters
behave as $q\gg1$ and $\tau_{0}^{-1}\approx\gamma S_{A}/4$. 

Thus the product of transformed waiting-time distributions in Eq.
(\ref{eq:product}) becomes\begin{equation}
\prod_{k=0}^{a-1}\underline{w}_{A-k,n-k}(z)=\frac{{n \choose a}{n+q \choose a}{n+2q \choose a}}{{n+\tau_{0}z \choose a}{n+q+\tau_{0}z \choose a}{n+2q+\tau_{0}z \choose a}},\end{equation}
where the Binomial coefficients are defined by means of the Gamma
function,\begin{equation}
{x \choose y}=\frac{\Gamma(x+1)}{\Gamma(x-y+1)\Gamma(y+1)}.\end{equation}
The Laplace transform of the no-escape probability is according to
Eq. (\ref{eq:no-escape-prob}) \begin{equation}
\underline{W}_{A-a,n-a}(z)=\frac{1-\underline{w}_{A-a,n-a}(z)}{z},\end{equation}
which can be written as the sum\begin{eqnarray}
 &  & \underline{W}_{A-a,n-a}(z)=\frac{\tau_{0}}{n-a+\tau_{0}z}\\
 &  & +\frac{\tau_{0}(n-a)}{\left(n-a+\tau_{0}z\right)\left(n-a+\tau_{0}z+q\right)}\nonumber \\
 &  & +\frac{\tau_{0}\left(n-a\right)\left(n-a+q\right)}{\left(n-a+\tau_{0}z\right)\left(n-a+\tau_{0}z+q\right)\left(n-a+\tau_{0}z+2q\right)}.\nonumber \end{eqnarray}

Thus the complete conditional probability (\ref{eq:product}) becomes
the sum of the three inverse Laplace transforms\begin{eqnarray}
 &  & P_{a}(\tau|n)=\frac{\tau_{0}e^{-n\tau/\tau_{0}}}{a+1}\mathcal{L}^{-1}\left[\frac{{n \choose a}{n+q \choose a}{n+2q \choose a}}{{\tau_{0}z \choose a+1}{q+\tau_{0}z \choose a}{2q+\tau_{0}z \choose a}}\right.\label{eq:laplace-expr}\\
 &  & \quad\left.+\frac{{n \choose a+1}{n+q \choose a}{n+2q \choose a}}{{\tau_{0}z \choose a+1}{q+\tau_{0}z \choose a+1}{2q+\tau_{0}z \choose a}}+\frac{{n \choose a+1}{n+q \choose a+1}{n+2q \choose a}}{{\tau_{0}z \choose a+1}{q+\tau_{0}z \choose a+1}{2q+\tau_{0}z \choose a+1}}\right].\nonumber \end{eqnarray}
Each Laplace transform results as a Meijer G function \citet{prudnikov},\begin{widetext}
\begin{eqnarray}
P_{a}(q,p|n) & = & (a!)^{3}{n+2q \choose a}(1-p)^{n}\left[{n \choose a}{n+q \choose a}G_{33}^{30}\left(1-p\left|\begin{array}{ccc}
1 & q+1 & 2q+1\\
-a & q-a+1 & 2q-a+1\end{array}\right.\right)\right.\nonumber \\
 &  & \quad\left.+(a+1){n \choose a+1}{n+q \choose a}G_{33}^{30}\left(1-p\left|\begin{array}{ccc}
1 & q+1 & 2q+1\\
-a & q-a & 2q-a+1\end{array}\right.\right)\right.\nonumber \\
 &  & \quad\left.+(a+1)^{2}{n \choose a+1}{n+q \choose a+1}G_{33}^{30}\left(1-p\left|\begin{array}{ccc}
1 & q+1 & 2q+1\\
-a & q-a & 2q-a\end{array}\right.\right)\right],\label{eq:P-Gfunc}\end{eqnarray}
\end{widetext}where we have introduced the parameter \begin{equation}
p=1-e^{-\tau/\tau_{0}}.\label{eq:xi-def}\end{equation}

\section{Discussion\label{sec:Discussion}}

\begin{figure}
\includegraphics[width=0.45\textwidth]{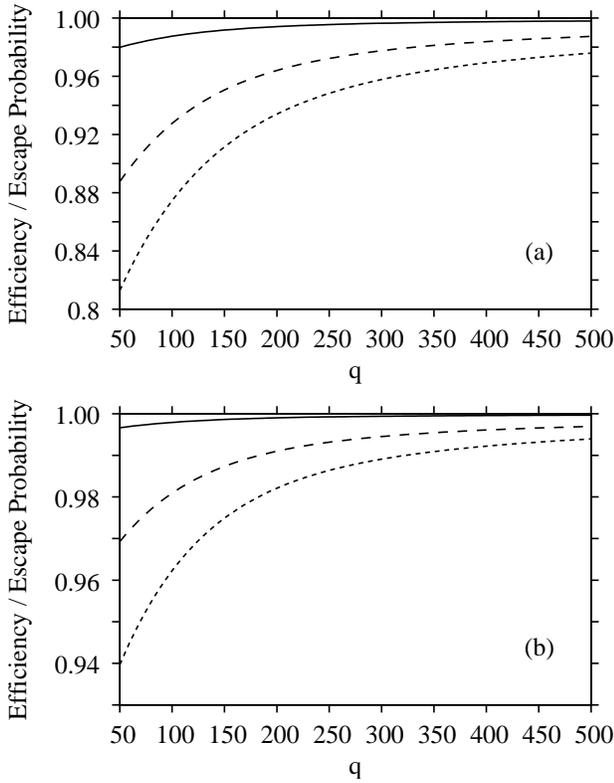}

\caption{Dependence of the scaled quantum efficiency $\eta_{D}/p$ on the saturation-dependent
parameter $q$ for $n=1$ (solid), $n=10$ (dashed), and $n=20$ (dotted)
photons. Atom-escape probabilities are $p=0.9$ (a) and $p=0.6$ (b).}

\label{fig:Dependence-of-the}
\end{figure}

The effective rate at which an atom leaves the trap is $\tau_{0}^{-1}\approx4|\Omega|^{2}A/\gamma$
so that the introduced parameter $p$, cf. Eq. (\ref{eq:xi-def}),
can be identified as the probability for an atom to escape from the
trap by the absorption of a photon. However, there is additionally
the large saturation-dependent parameter $q\approx2/S_{A}$, cf. Eq.
(\ref{eq:q-def}), that may change the shape of the conditional probability
(\ref{eq:P-Gfunc}). It is thus not obvious how these two parameters
can be merged into a possibly existing single parameter, such as the
quantum efficiency. 

However, such a quantum efficiency may be defined phenomenologically,
demanding that the average atom count conditioned on the presence
of $n$ photons reads \begin{equation}
\bar{a}_{n}=\eta_{D}n.\label{eq:etaD-def}\end{equation}
Thus, only the fraction $\eta_{D}$ of the $n$ photons leads on average
to $\bar{a}_{n}$ escaping atoms. In general, the above relation is
not necessarily linear, as the efficiency may be a function of the
photon number, revealing a nonlinear relation between incoming photon
and escaping atom numbers. For our specific case, such a phenomenological
quantum efficiency additionally depends on the atom-escape probability
$p$ and the saturation-dependent coefficient $q$, thus we obtain
\begin{equation}
\eta_{D}=\eta_{D}(q,p,n).\end{equation}

In Fig. \ref{fig:Dependence-of-the} (a) this dependence is shown
for an atom-escape probability $p=0.9$ in dependence of the parameter
$q$ for $n=1$ (solid), $n=10$ (dashed), and $n=20$ (dotted) photons.
It can be observed that for large values of $q$ all these curves
converge to the value of the atom-escape probability, i.e. \begin{equation}
\lim_{q\to\infty}\eta_{D}(q,p,n)=p.\end{equation}
The same behavior is also observed for a lower escape probability
$p=0.6$, see Fig. \ref{fig:Dependence-of-the} (b). Thus for sufficiently
low saturation $S_{A}$, i.e. for sufficiently large $q$, a linear
regime is attained, where the quantum efficiency becomes independent
of the incident photon number $n$ and coincides with the atom-escape
probability $p$. 

\begin{figure}
\includegraphics[width=0.48\textwidth]{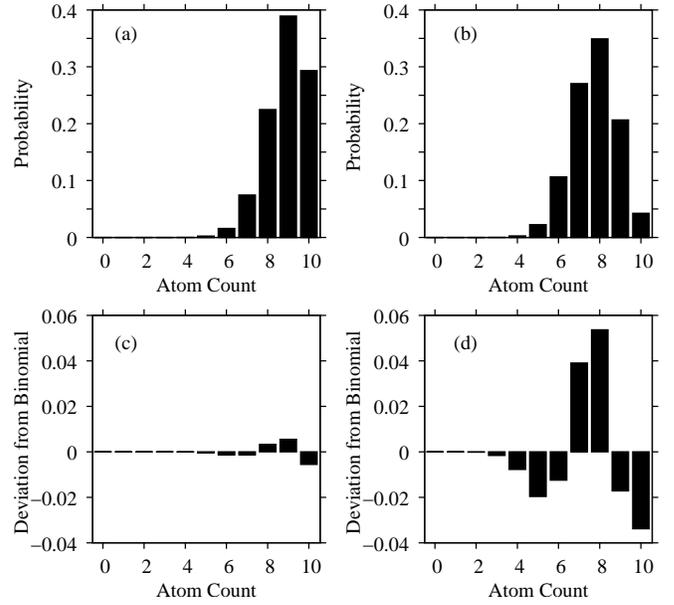}

\caption{Conditional atom-count statistics $P_{a}(q,p|n)$ for $n=10$ photons,
$p=0.9$, $q=100$ (a) and $q=10$ (b). The lower part shows the deviation
from the corresponding Binomial statistics with quantum efficiency
$\eta_{D}=0.8862$ for $q=100$ (c) and $\eta_{D}=0.7730$ for $q=10$
(d).}

\label{fig:counts-10-09-100-10}
\end{figure}

Does the statistics become then identical to that known from the Mandel
counting formula? To answer this question, we proceed as follows:
Given that by use of Eq. (\ref{eq:etaD-def}) we may identify a quantum
efficiency $\eta_{D}$ for each conditional atom-count statistics
$P_{a}(q,p|n)$, we may compare it with the corresponding conditional
count-statistics that would correspond to the Mandel formula of photo-detection
(\ref{eq:mandel}). The latter is given by the Binomial statistics,\begin{equation}
P_{a}^{(M)}(\eta_{D}|n)={n \choose a}\eta_{D}^{a}(1-\eta_{D})^{n-a}.\label{eq:binomial}\end{equation}
In Fig. \ref{fig:counts-10-09-100-10} (a) the atom-count statistics
is shown for an incident light field with $n=10$ photons, with an
atom-escape probability $p=0.9$ and $q=100$. From part (c) it can
be seen that the deviation from the Binomial statistics of corresponding
quantum efficiency $\eta_{D}=0.8862$ is not too large. However, this
changes if the saturation-dependent coefficient is lowered to $q=10$,
see the atom-count statistics shown in Fig. \ref{fig:counts-10-09-100-10}
(b). Now the deviation from a Binomial statistics with corresponding
quantum efficiency $\eta_{D}=0.7730$ becomes rather large, cf. Fig.
\ref{fig:counts-10-09-100-10} (d). In fact, the statistics becomes
narrower as compared to the Binomial form.

\begin{figure}
\includegraphics[width=0.48\textwidth]{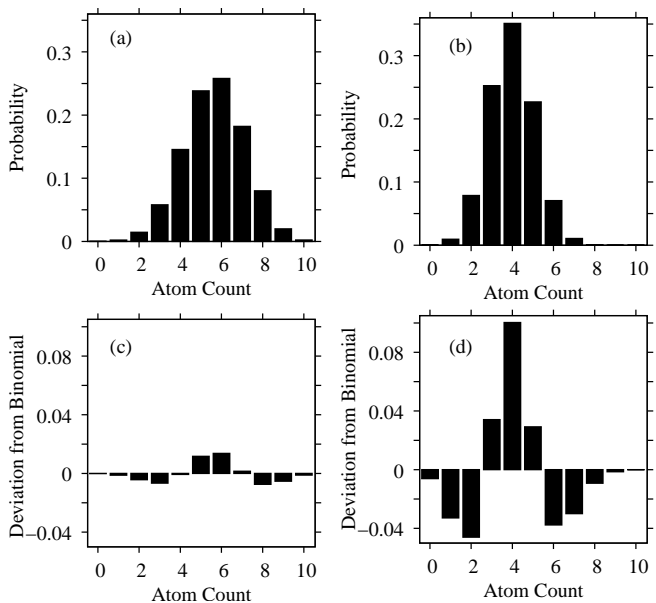}

\caption{Conditional atom-count statistics $P_{a}(q,p|n)$ for $n=10$ photons,
$p=0.6$, $q=100$ (a) and $q=10$ (b). The lower part shows the deviation
from the corresponding Binomial statistics with quantum efficiency
$\eta_{D}=0.5639$ for $q=100$ (c) and $\eta_{D}=0.3968$ for $q=10$
(d).}

\label{fig:counts-10-06-100-10}
\end{figure}

Whereas these two cases used a rather large atom-escape probability
of $p=0.9$, for a lower value such as $p=0.6$ already for $q=100$
a somewhat larger deviation from the Binomial statistics can be observed
in Fig. \ref{fig:counts-10-06-100-10} (c). However, the general trend
of larger deviation for smaller values of $q$ is confirmed.

We may thus conclude that for a given value of $p$ in the limit $q\to\infty$
we approach the form of a Binomial statistics, agreeing with the Mandel
formula. This limit corresponds to $S_{A}\to0$, i.e. to extremely
low saturation. For such low saturation the atomic gas absorbs photons
one by one, making the appearance of bunches of simultaneously escaping
atoms highly improbable. As in this limit, at a given instant of time,
no more than one escaped atom may be detected, the indistinguishability
of the bosonic atoms can be safely disregarded. Thus the Mandel formula
is reproduced, as given by its perturbative derivation assuming distinguishable
photo-electron emission centers. 

For larger saturation, however, it becomes more probable that several
photons are absorbed simultaneously, generating thus bunches of escaping
atoms. In this regime the indistinguishability of the bosonic atoms
becomes relevant and large deviations from the Mandel counting formula
are observed. They manifest themselves by a reduced fluctuation of
atom counts as compared to the Binomial form. This noise reduction
leads to an enhanced photon-number resolution as compared to a conventional
photodetector at equal detection efficiency, which constitutes an
important advantage of a micro-trap based detector.

It should be noted that the detection efficiency $\eta_{D}$ determines
the mapping photon to atom numbers to all orders of moments for a
conventional photodetector. In our case, however, it only connects
the mean photon and atom-count numbers. The shape and width of the
atom-count statistics additionally depends on a second parameter,
e.g. the saturation, which allows for a difference from the Binomial
statistics. 

The present model assumes an ideal Bose gas in the limit of zero temperature.
However, the inclusion of atomic collisions can be consistently made
by identifying the lowest trap mode as the solution of the Gross--Pitaevskii
equation. In Thomas--Fermi approximation the only modification will
be then the increased (decreased) size of the ground-state mode for
repulsive (attractive) scattering, which will modify the time of flight
of excited atoms to escape from the trap. Thus repulsive (attractive)
scattering will decrease (increase) the atom-loss rate $\gamma$.

The treatment of finite temperature (but below the condensation temperature)
requires the departure from the single-mode approximation. However,
for low saturation as discussed here, the inclusion of non-condensed
modes will not lead to a mixing of condensed and non-condensed atoms
by photon absorption: Photon absorption will excite both initially
condensed as well as non-condensed atoms to immediately leave the
trap. Thus they do not return to the electronic ground state, preventing
the mixing. The small fraction of non-condensed atoms at finite temperature
will represent distinguishable emission centers that will contribute
as a small Binomial admixture to the atom-count statistics. With increasing
temperature this admixture will deterioate the enhancement of photon-number
resolution made by the condensed atoms, as we expect from the transition
beyond the condensation temperature to a gas of distinguishable atoms.

\section{Summary and Conclusions\label{sec:Summary-and-Conclusions}}

In summary we have introduced a model for a photo detector using a
Bose-condensed gas trapped on an atom chip. By use of an atomic pseudo-spin
approximation based on large atom numbers, the photon absorption and
atom-escape dynamics could be approximated for the over damped case
where Rabi oscillations are suppressed. A quantum-trajectory method
then led us to a counting formula involving a sum of three Meijer
G functions. 

The conditional count statistics has been shown to depend on a saturation-dependent
coefficient $q$ and an atom-escape probability $p$, to which the
quantum efficiency of the detector converges in the limit of extremely
low saturation, $q\to\infty$. In this limit the differences from
a Binomial statistics vanish so that our result coincides with the
Mandel counting formula. However, for realistic values of the saturation
substantial deviations can be observed as a reduced atom-count fluctuations.

An open question is yet the behavior of our model detector for saturation
$S_{A}\geq1$, where full Rabi oscillations start to occur. We assume
that in this regime we may expect even larger and dramatic deviations
from the Mandel formula. Then coherent processes must be included,
that describe the temporary storage of photon energy in the atomic
gas in the absence of correspondingly escaping atoms. This is left
for a future investigation.

\begin{acknowledgments}
The authors acknowledge support by FONDECYT project no. 7070220 and
S.W. acknowledges support by FONDECYT project no. 1051072.
\end{acknowledgments}
\appendix

\section{Transition amplitude\label{sec:Transition-amplitude}}

The transition amplitude can be simplified to\begin{eqnarray*}
 &  & \Psi_{n-1,n}^{A,n}(\tau)=\sqrt{\gamma}e^{-\gamma n\tau/4}\\
 &  & \quad\times\langle A,n,n-1|e^{i\tau|(2|\Omega|\sqrt{M_{A,n}}\hat{L}_{x}-i\gamma\hat{L}_{z}/2)}|A,n,n\rangle,\end{eqnarray*}
where $M_{A,n}=A-(n-1)/2$. The above matrix element can be calculated
as\[
\langle A,n,n-1|e^{i(\alpha\hat{L}_{z}+\beta\hat{L}_{x})}|A,n,n\rangle=\sqrt{n}z^{-n/2}\zeta^{n-1}(1+z)^{n-1},\]
where \begin{eqnarray*}
\zeta & = & \frac{i\alpha\sin(\delta/2)}{\delta\cos(\delta/2)-i\beta\sin(\delta/2)},\\
z & = & -\frac{\delta^{2}}{\alpha^{2}\sin^{2}(\delta/2)},\\
\delta & = & \sqrt{\alpha^{2}+\beta^{2}},\end{eqnarray*}
using $\alpha=2|\Omega|\tau\sqrt{M_{A,n}}$ and $\beta=-i\gamma\tau/2$
for the resonant case ($\Delta=0$). 

Using the explicit values for $\alpha$ and $\beta$ we obtain the
explicit expressions for the above parameters with $\delta=i\delta^{\prime}$,\begin{eqnarray*}
\zeta & = & \frac{i\sqrt{S_{A,n}}\sinh(\delta^{\prime}/2)}{\sqrt{1-S_{A,n}}\cosh(\delta^{\prime}/2)-\sinh(\delta^{\prime}/2)},\\
z & = & \frac{S_{A,n}-1}{S_{A,n}\sinh^{2}(\delta^{\prime}/2)},\\
\delta^{\prime} & = & \frac{\gamma\tau}{2}\sqrt{1-S_{A,n}},\end{eqnarray*}
where we defined the saturation parameter, \begin{eqnarray}
S_{A,N} & = & \frac{4|\Omega|^{2}M_{A,N}}{\left(\frac{\gamma}{2}\right)^{2}}.\label{eq:saturation1}\end{eqnarray}

In the over damped regime, $S_{A,n}<1$, the transition amplitude
becomes thus\begin{eqnarray*}
\Psi_{n-1,n}^{A,n}(\tau) & = & -i\sqrt{n\gamma}e^{-\gamma n\tau/4}(-1)^{n-1}\\
 & \times & \left[\sqrt{1-S_{A,n}}\cosh(\delta^{\prime}/2)+\sinh(\delta^{\prime}/2)\right]^{n-1}\\
 & \times & \left[\frac{1}{(1-S_{A,n})^{n/2}}\right]\sqrt{S}\sinh(\delta^{\prime}/2).\end{eqnarray*}
In the limit $S_{A,n}\ll1$ this expression reduces in lowest-order
approximation to\begin{eqnarray}
\Psi_{n-1,n}^{A,n}(\tau) & = & i(-1)^{n}\sqrt{S_{A,n}n\gamma}e^{-\frac{\gamma\tau}{4}[n-(n-1)\sqrt{1-S_{A,n}}]}\nonumber \\
 &  & \times\sinh\left(\frac{\gamma\tau}{4}\sqrt{1-S_{A,n}}\right).\label{eq:psi-approx}\end{eqnarray}


\begin{thebibliography}{10}
\bibitem{mandel2} See L.~Mandel and E.~Wolf, \emph{Optical Coherence
and Quantum Optics} (Cambridge University Press, 1995) and references
therein.

\bibitem{kelley}P.L. Kelley and W.H. Kleiner, Phys. Rev. \textbf{136},
A316 (1964); R.J. Glauber, \emph{Quantum Optics and Electronics} (Les
Houches Summer School of Theoretical Physics, University of Grenoble),
eds C. DeWitt, A. Blandin, and C. Cohen-Tannoudji (Gordon and Breach,
New York, 1965), p.53.

\bibitem{mandel1}L. Mandel, Proc. Phys. Soc. (London) \textbf{72},
1037 (1958); \textbf{74}, 233 (1959); \emph{ibid.}, Prog. Opt. \textbf{2},
181 (1963). 

\bibitem{mandel3}L. Mandel, E.C.G. Sudarshan, and E. Wolf, Proc.
Phys. Soc. (London) \textbf{84}, 435 (1964); W.E. Lamb Jr. and M.O.
Scully, in \emph{Polarization: Matiere et Rayonnement} (Presses Universitaires
de France, Paris, 1969). 

\bibitem{bec}M.H. Anderson, J.R. Ensher, M.R. Matthews, C.E. Wieman,
and E.A. Cornell, Science \textbf{269}, 198 (1995); C.C. Bradley,
C.A. Sackett, J.J. Tollett, and R.G. Hulet, Phys. Rev. Lett. \textbf{75},
1687 (1995); K.B. Davies, M.-O. Mewes, M.R. Andrews, N.J. van Druten,
D.S. Durfee, D.M. Kurn, and W. Ketterle, \emph{ibid.} \textbf{75},
3969 (1995); 

\bibitem{bec-on-atom-chip}H. Ott, J. Fortágh, G. Schlotterbeck, A.
Grossmann, and C. Zimmermann, Phys. Rev. Lett. \textbf{87}, 230401
(2001); W. Hänsel, J. Reichel, P. Hommelhoff, and T.W. Hänsch, Phys.
Rev. A \textbf{64}, 063607 (2001).

\bibitem{atom-chip}See also J. Fortágh and C. Zimmermann, Rev. Mod.
Phys. \textbf{79}, 235 (2007) and references therein.

\bibitem{dicke} R.H. Dicke, Phys. Rev. \textbf{93}, 99 (1954).

\bibitem{chumakov1}M. Kozierowski, A.A. Mamedov, and S.M. Chumakov,
Phys. Rev. A \textbf{42}, 1762 (1990).

\bibitem{chumakov2}S.M. Chumakov, A.B. Klimov, and M. Kozierowski,
\emph{From the Jaynes-Cummings model to collective interactions} in
\emph{Theory of nonclassical states of light}, eds V.V. Dodonov and
V.I. Man'ko (Taylor \& Francis, London, 2002) p. 313.

\bibitem{qt1}G.C. Hegerfeldt and T.S. Wilser, in \emph{Classical
and quantum system}s, Proc. of the \emph{II. International Wigner
Symposium}, 1991 (World Scientific, Singapore, 1992), p. 104; C.W.
Gardiner, A.S. Parkins, and P. Zoller, Phys. Rev. A \textbf{46}, 4363
(1992); J. Dalibard, Y. Castin, and K. Molmer, Phys. Rev. Lett. \textbf{68},
580 (1992).

\bibitem{qt2}H.J. Carmichael, \emph{An open systems approach to quantum
optics}, vol. m18 of \emph{Lecture notes in physics} (Springer Verlag,
Berlin, Heidelberg, New York, 1993).

\bibitem{qt3}K. Molmer, Y. Castin, and J. Dalibard, J. Opt. Soc.
Am. B \textbf{10}, 524 (1993); G.C. Hegerfeldt and M.B. Plenio, Quant.
Opt. \textbf{6}, 15 (1994); B.M. Garraway and P.L. Knight, Phys. Rev.
A \textbf{50}, 2548 (1994); M.B. Plenio and P.L. Knight, Rev. Mod.
Phys. \textbf{70}, 101 (1998).

\bibitem{prudnikov}A.P. Prudnikov, Yu. A. Brychkov, and O.I. Marichev,
\emph{Integrals and Series}, vol. 4 (Gordon and Breach Science Publishers,
New York, 1992), p. 578.
\end{thebibliography}
\end{document}